\documentclass[10pt,a4papper]{article}
\usepackage[utf8]{inputenc}
\usepackage{indentfirst}
\usepackage[]{graphicx}
\usepackage{geometry} 
\begin{document}

\sloppypar
	
\title{Processes $\tau^- \to \pi^- \pi^0 \nu_{\tau}$ and $e^{+}e^{-} \to \pi^{+} \pi^{-}$ 
in the chiral NJL model with final state interactions taken into account}

\author{M. K. Volkov\footnote{volkov@theor.jinr.ru}, 
A. B. Arbuzov\footnote{arbuzov@theor.jinr.ru}, 
A. A. Pivovarov\footnote{tex$\_$k@mail.ru}\\
{\it Bogoliubov Laboratory of Theoretical Physics, Joint Institute for Nuclear Research,\/} \\ 
{\it Dubna, 141980, Russia}}
\date{}

\maketitle
	
\begin{abstract}
The processes $\tau^- \to \pi^- \pi^0 \nu_{\tau}$ and $e^{+}e^{-} \to \pi^{+} \pi^{-}$
are considered within the chiral Nambu--Jona-Lasinio model with taking into account 
the pion final state interactions beyond the leading $1/N_c$ approximation. 
The contribution of the loop correction caused by a $\rho$ meson exchange between 
the pions, which gives the main contribution in the P-wave channel, is caclulated. 
The results for both processes are in a satisfactory agreement with experimental data.
\end{abstract}

\section{Introduction}

The chiral Nambu--Jona-Lasinio (NJL) model~\cite{Nambu:1961tp,Ebert:1982pk,Volkov:1986zb,Ebert:1985kz} 
and its extended version~\cite{Volkov:1996br,Volkov:1996fk,Volkov:1997dd}, which takes into account 
the first radially excited meson states, allowed to describe many $\tau$ lepton decay modes and 
the processes of meson production in colliding $e^{+}e^{-}$ beams at low energies 
in a satisfactory agreement with experimental
data~\cite{Volkov:2016umo,Volkov:2017arr,Volkov:2018zlr,Volkov:2019cja,Volkov:2019jug,Volkov:2019izp,Volkov:2019yli}. 
However the description of such important processes as the decay $\tau^- \to \pi^- \pi^0 \nu_{\tau}$ 
and the production of a pion pair in $e^{+}e^{-}$ annihilation at low energies was a problem for the model
especially taking into account the corresponding high experimental accuracy. 
Here we consider the possibility to take into account the pion interaction
in the final states of these processes. We suggest to include contributions of
Feynman diagrams with $\rho$ meson exchange between the pions, which are relevant for 
the P-wave final state. 
These amplitudes are in a higher order of the $1/N_c$ expansion compared to 
the standard NJL model approximation. 
	
The process $\tau^- \to \pi^- \pi^0 \nu_{\tau}$ is the most probable $\tau$ lepton decay mode. 
Its partial width is measured with high precision: 
$\mathrm{Br}(\tau \to \pi \pi \nu_{\tau}) = (25.49 \pm 0.09)\%$~\cite{Tanabashi:2018oca}. 
In theoretical works, an agreement with experimental data is usually achieved by using phenomenological 
parametrizations of the pion form factor fitted by the experimental 
data~\cite{Kuhn:1990ad,Bartos:2017oam,Miranda:2018cpf,Dai:2018thd} without analysis of the corresponding 
physical picture. The same situation takes the place for the process $e^{+}e^{-} \to \pi^{+} \pi^{-}$. 
Its hadron current is related to the corresponding current in the $\tau$ lepton decay by an isospin transformation 
in the framework of the vector current conservation hypothesis. Therefore, it is important to have a concerted 
theoretical description of both processes. Besides that, the problem to concert the low energy processes $e^{+}e^{-}$ 
annihilation to hadrons and the corresponding modes of the $\tau$ lepton decays is discussed in the different papers, 
see the work~\cite{Benayoun:2015hzf} and references therein. On the other hand, the precision of the experimental 
results on these processes increases permanently~\cite{Achasov:2005rg,Schael:2005am,Fujikawa:2008ma}. 
The importance of a detailed understanding of these processes is due to the fact that they give a significant 
contribution to the definition of the hadron vacuum polarisation~\cite{Actis:2010gg}. In the process $e^{+}e^{-}$ 
annihilation to pions, it is also important to describe the mixing of the $\rho$ and $\omega$ mesons which 
is sensitive to the difference of the current masses of $u$ and $d$ quarks.
	
The pion interaction in the final state has been studied theoretically within a number of approaches. The most general 
approach is the method of dispersion relations. The interactions of the light mesons pairs $\pi\pi$, $\pi K$, and $KK$ 
with considering of phase shifts were successfully described in the framework of this approach~\cite{Dax:2018rvs}. 
It is important to take into account these interactions for the $\pi\pi$ and $KK$ scattering description at low energies. 
The considered energy range is also in the applicability region of the 
Chiral Perturbation Theory (ChPT)~\cite{Weinberg:1978kz,Gasser:1983yg}. 
In particular, in the framework of the so-called unitarized ChPT, the interactions in the final state were researched with 
the Inverse Amplitude Method~\cite{Truong:1991gv,GomezNicola:2007qj}. The direct application of these methods 
in combination with the NJL model is impossible due to the inconsistency of the corresponding approaches. 
On the other hand, the meson loop corrections have been used in the framework of the NJL model in description 
of the series processes~\cite{Ebert:1996pc,Volkov:2009pc}. However, the considered meson loops were in  
the leading approximation of the model in $1/N_c$.
	
This paper is organized in the following way. In Chapter~\ref{sect:NJL}, the relevant Lagrangian of the standard NJL 
model is given. In the next two Chapters,~\ref{sect:ee} and \ref{sect:tau}, the processes of $e^{+}e^{-}$ annihilation 
into two pions and of $\tau$ lepton decay are considered. Chapter~\ref{sect:concl} is devoted to discussion of results.

\section{The Lagrangian of the standard NJL model} \label{sect:NJL}

The relevant fragment of the quark-meson interaction Lagrangian of the NJL model with the vertices 
we need has the form~\cite{Volkov:1986zb}:
\begin{eqnarray}
\label{Lagrangian}
\Delta \mathcal{L}_{int} & = &
\bar{q} \left[ \frac{g_{\rho}}{2} \gamma^{\mu} \left(\tau_{3} \rho_{\mu}^{0} + \tau_{0} \omega_{\mu}\right) 
+ i g_{\pi} \gamma^{5} \tau_{3} \pi^{0} + \frac{g_{\rho}}{2} \gamma^{\mu} \sum_{j = \pm} \tau_{j} \rho_{\mu}^{j} 
+ i g_{\pi} \gamma^{5} \sum_{j = \pm} \tau_{j} \pi^{j} \right]q,
\end{eqnarray}
where $q$ and $\bar{q}$ are the $SU(2)$ doublets of the $u$ and $d$ quark fields with the constituent 
masses $m_{u} \approx m_{d} = m = 280$~MeV; $\tau_{3}$ is the Pauli matrix; $\tau_{\pm}$ are the linear 
combinations of the Pauli matrices; $\tau_{0}$ is the unit matrix.
	
The coupling constants are
\begin{eqnarray}
\label{Couplings}
g_{\rho} =  \left(\frac{2}{3}I_{2}\right)^{-1/2}, \quad g_{\pi} =  \left(\frac{4}{Z_{\pi}}I_{2}\right)^{-1/2},
\end{eqnarray}
where
\begin{eqnarray}
Z_{\pi} = \left[1 - 6\frac{m^{2}}{M_{a_{1}}^{2}}\right]^{-1}
\end{eqnarray}
is the additional renormalization constant appearing due to $\pi - a_{1}$ transitions, 
$M_{a_{1}} = 1230 \pm 40$~MeV~\cite{Tanabashi:2018oca} is the mass of the $a_1$ axial vector meson.
	
The integrals appearing as a result of the Lagrangian renormalization read
\begin{eqnarray}
\label{int}
I_{2} =
-i\frac{N_{c}}{(2\pi)^{4}}\int\frac{\Theta(\Lambda^{2} + k^2)}{(m^{2} - k^2)^{2}} \mathrm{d}^{4}k 
= \frac{N_{c}}{(4\pi)^{2}}\left[\ln\left(\frac{\Lambda^{2}}{m^{2}} + 1\right) 
- \left(1 + \frac{m^{2}}{\Lambda^{2}}\right)^{-1}\right],
\end{eqnarray}
where $\Lambda = 1250$~MeV is the cut-off parameter of the quark momentum in loop integrals, 
$N_{c} = 3$ is the number of colors.

\section{The process $e^{+}e^{-} \to \pi^{+} \pi^{-}$} \label{sect:ee}

\begin{figure}[ht]
\center{\includegraphics[scale = 0.6]{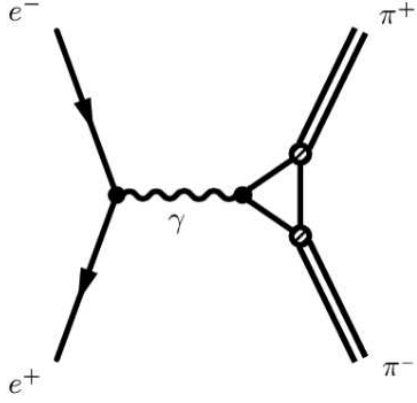}}
\caption{The contact diagram of the process $e^{+}e^{-} \to \pi^{+} \pi^{-}$.}
\label{eeContact}
\end{figure}

\begin{figure}[ht]
\center{\includegraphics[scale = 0.8]{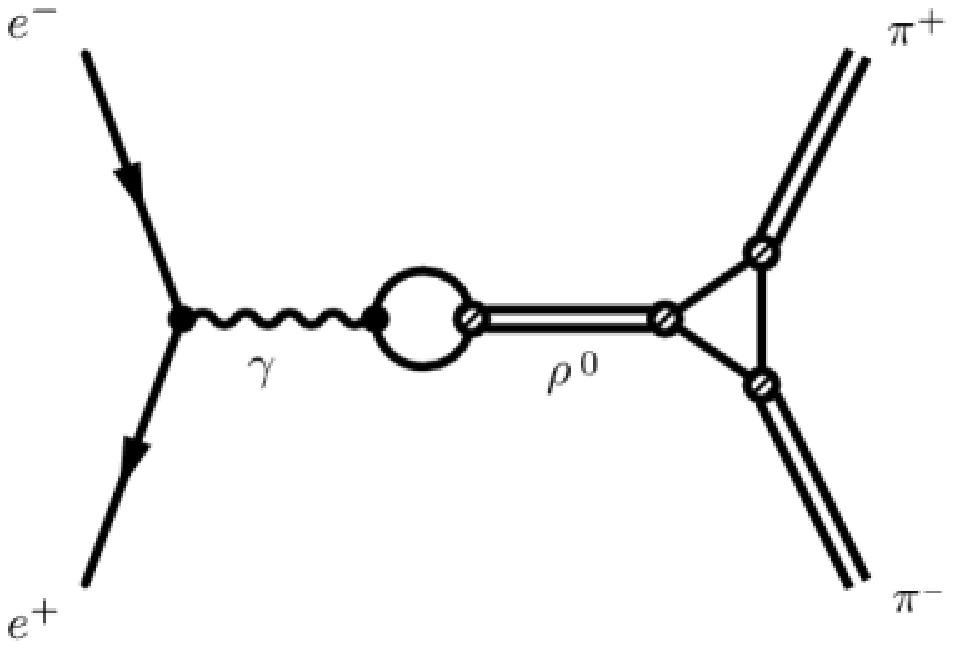}}
\caption{The diagram of the process $e^{+}e^{-} \to \pi^{+} \pi^{-}$ with an intermediate $\rho$ meson.}
\label{eeInterm}
\end{figure}

\begin{figure}[ht]
\center{\includegraphics[scale = 0.7]{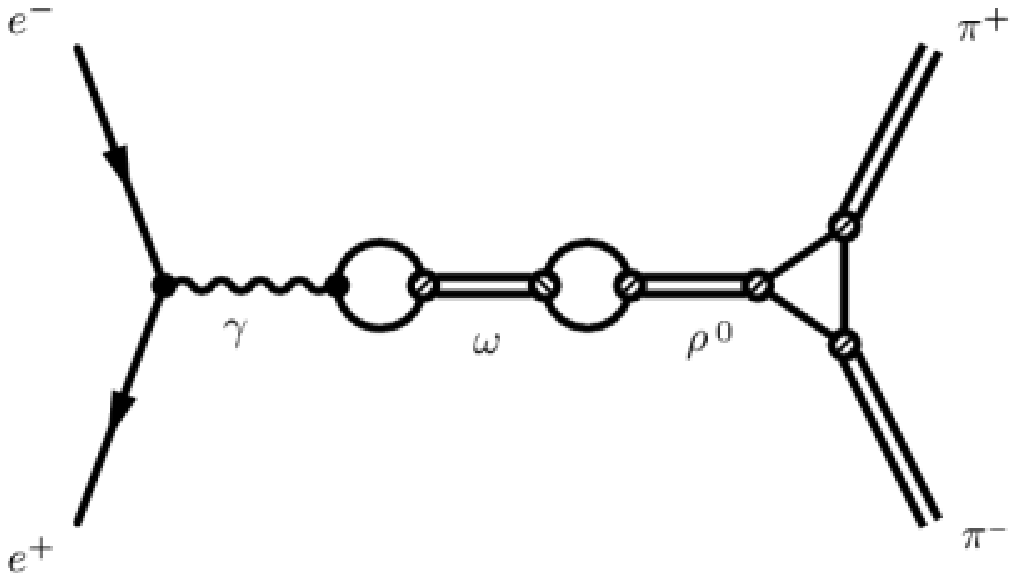}}
\caption{The diagram of the process $e^{+}e^{-} \to \pi^{+} \pi^{-}$ with an intermediate $\omega$ meson.}
\label{eeOmega}
\end{figure}

The diagrams of the process $e^{+}e^{-} \to \pi^{+} \pi^{-}$ without considering interaction 
in the final state are shown in Figs.~\ref{eeContact},~\ref{eeInterm},~\ref{eeOmega}.

\begin{figure}[ht]
\center{\includegraphics[scale = 0.6]{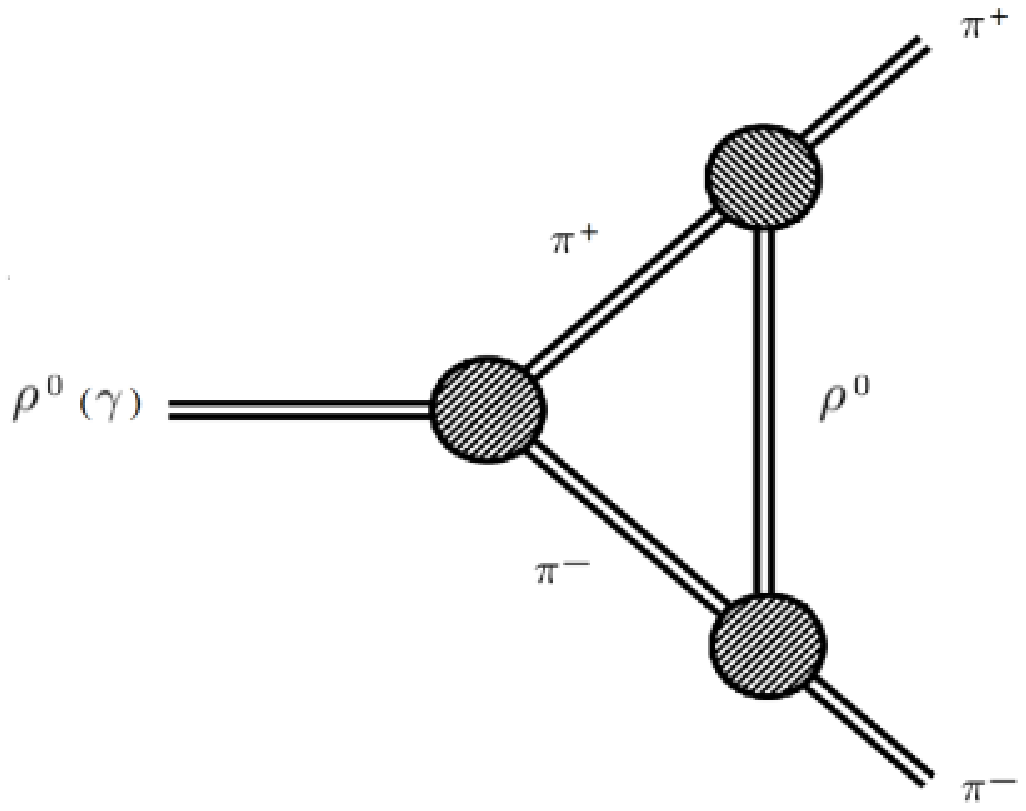}}
\caption{The pion final state interaction in the case of $e^{+}e^{-}$ annihilation.}
\label{rho0_gamma_pipi}
\end{figure}

In this process, only the P-wave takes place. The box diagram of the four pion contact interaction and the pion 
interaction via the scalar meson represent the S-wave and do not contribute to this process. The pion interaction 
by means of annihilation into $\rho$ meson is taken into account in the decay width in the Breit-Wigner propagator 
denominator of the intermediate $\rho$ meson. That is why it is enough to take into account the $\rho$ meson 
exchange between the pions for considering the pion interaction in the final state. As a result, the meson 
triangle appears, see Fig.~\ref{rho0_gamma_pipi}.

Every vertex of this triangle can be written via the amplitude of the decay $\rho^{0} \to \pi^{+} \pi^{-}$ 
which can be obtained with the Lagrangian~(\ref{Lagrangian}):
\begin{eqnarray}
M(\rho^{0} \to \pi^{+} \pi^{-}) = g_{\rho} e_{\mu}(q) \left(p_{+} - p_{-}\right)^{\mu},
\end{eqnarray}
where $q$ is the $\rho$ meson momentum, $p_{+}$ and $p_{-}$ are pion momenta, $e_{\mu}(q)$ is 
the polarisation vector of the $\rho$ meson.
	
By using this amplitude one can obtain the vertex of the meson Lagrangian describing 
the interaction of a $\rho$ meson with pions:
\begin{eqnarray}
\mathcal{L} = -i g_{\rho} \rho_{\mu}^{0} \left(\pi^{+}\partial^{\mu}\pi^{-} 
- \pi^{-}\partial^{\mu}\pi^{+}\right).
\end{eqnarray}
	
The triangle meson loop can be constructed from these vertices. It leads to the integral:
\begin{eqnarray}
g_{\rho}^{3} \int \frac{\left(k - 2p_{-}\right)^{\lambda}\left(k + 2p_{+}\right)^{\nu}
\left(2k + p_{+} - p_{-}\right)^{\mu} \left(g_{\nu\lambda} 
- \frac{k_{\nu}k_{\lambda}}{M_{\rho}^{2}}\right)}{\left[k^{2} - M_{\rho}^{2}\right]
\left[(k - p_{-})^{2} - M_{\pi}^{2}\right]\left[(k + p_{+})^{2} - M_{\pi}^{2}\right]} 
\frac{d^{4}k}{(2\pi)^{4}}.
\end{eqnarray}
By the series expansion of this integral in powers of the external momenta and keeping only 
divergent terms (similarly to the method applying in the NJL model), we obtain
\begin{eqnarray}
i g_{\rho}^{3} T_{\rho\pi\pi} \left(p_{+} - p_{-}\right)^{\mu},
\end{eqnarray}
where 
\begin{eqnarray}
T_{\rho\pi\pi} = \frac{I_{1M}}{M_{\rho}^{2}} + I_{2M},
\end{eqnarray}
$q = p_{+} + p_{-}$; $I_{1M}$ and $I_{2M}$ are quadratically and logarithmically divergent 
integrals, respectively:
\begin{eqnarray}
I_{2M} & = &
\frac{-i}{(2\pi)^{4}}\int\frac{\Theta(\Lambda_{M}^{2} + k^2)}{(M_{\rho}^{2} - k^2)(M_{\pi}^{2} - k^2)} \mathrm{d}^{4}k 
= \frac{1}{(4\pi)^{2}}\frac{1}{M_{\rho}^{2} - M_{\pi}^{2}}\left[M_{\rho}^{2}
\ln\left(\frac{\Lambda_{M}^{2}}{M_{\rho}^{2}} + 1\right) 
- M_{\pi}^{2}\ln\left(\frac{\Lambda_{M}^{2}}{M_{\pi}^{2}} + 1\right)\right], 
\nonumber\\
I_{1M} & = & \frac{-i}{(2\pi)^{4}}\int\frac{\Theta(\Lambda_{M}^{2} + k^2)}{(M_{\rho}^{2} - k^2)} \mathrm{d}^{4}k 
= \frac{1}{(4\pi)^{2}} \left[\Lambda_{M}^{2} - M_{\rho}^{2}\ln\left(\frac{\Lambda_{M}^{2}}{M_{\rho}^{2}} + 1\right)\right],
\end{eqnarray}
where $\Lambda_{M}$ is the cut-off parameter for the meson momentum in the loop.
	
Then the amplitude of the $e^{+}e^{-}$ annihilation process into the mesons takes the form
\begin{eqnarray}
M(e^{+}e^{-} \to \pi \pi) & = & -\frac{4 \pi \alpha_{em}}{s} 
\left[1 + \frac{s}{M_{\rho}^{2} - s - i \sqrt{s}\Gamma_{\rho}} 
+ \frac{s^{2}}{9} \frac{g_{\rho}^{2}\left[I_{2}(u) - I_{2}(d)\right]}{\left[M_{\rho}^{2} 
- s - i \sqrt{s}\Gamma_{\rho}\right]\left[M_{\omega}^{2} - s - i \sqrt{s}\Gamma_{\omega}\right]}\right] 
\times \nonumber\\
&& \times \left\{1 + g_{\rho}^{2}T_{\rho\pi\pi}\right\} 
L_{\mu}^{em} (p_{+} - p_{-})^{\mu}, 
\end{eqnarray}
where $\alpha_{em} \approx 1/137$ is the fine structure constant, 
$s = (p_{+} + p_{-})^{2}$, $L_{\mu}^{em} = \bar{e}^{-}\gamma_{\mu}e^{-}$ is the electromagnetic lepton current.
	
The third term in the first squared brackets describes the contribution of the diagram with the intermediate 
$\omega$ meson. In this term, the expression $I_{2}(u) - I_{2}(d)$ appears. Here $I_{2}(u)$ and $I_{2}(d)$ 
are the integrals of the form~(\ref{int}) with the masses of $u$ and $d$ quarks, respectively. In that case, 
the difference of these quark masses should not be neglected. We have used the value $m_{d} - m_{u} = 4$~MeV, 
obtained in the framework of the NJL model from the description the decay $\omega\to \pi\pi$~\cite{Volkov:1986zb}.
	
\begin{figure}[ht]
\center{\includegraphics[scale = 0.5]{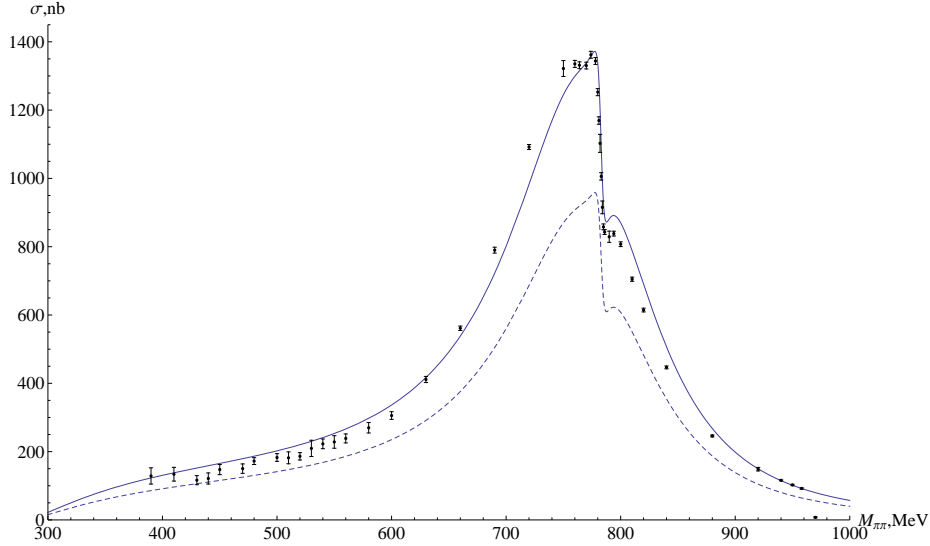}}
\caption{The dependence of the cross-section of the process $e^{+}e^{-} \to \pi^+ \pi^-$ on the colliding 
leptons energy in the center of mass system. The experimental points are taken from the paper~\cite{Achasov:2005rg}. 
The solid line describes the case with interaction in the final state and the dashed line describes the case 
without this interaction.}
\label{CrossSection}
\end{figure}

By experimentally known dependence of the cross-section of this process on the colliding leptons energy, 
one can fix the cut-off parameter of the meson momentum in the triangle $\Lambda_{M} = 740$~MeV. 
The dependence obtained for this cut-off parameter is shown in Fig.~\ref{CrossSection}.
By comparison with the experimental data~\cite{Achasov:2005rg} one can see that the interaction in the final 
state is very important near the peak corresponding to $\rho$ meson resonance.

\section{The process $\tau^- \to \pi^- \pi^0 \nu_{\tau}$} \label{sect:tau}

\begin{figure}[ht]
\center{\includegraphics[scale = 0.6]{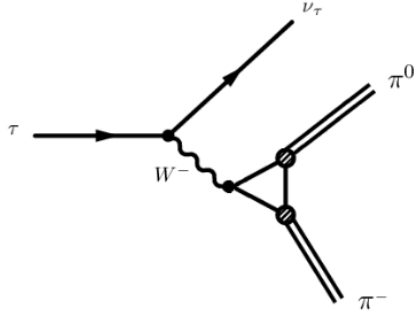}}
\caption{The contact diagram of the process $\tau^- \to \pi^- \pi^0 \nu_{\tau}$.}
\label{tauContact}
\end{figure}

\begin{figure}[ht]
\center{\includegraphics[scale = 0.7]{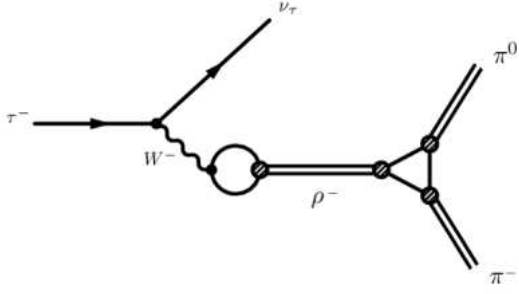}}
\caption{The diagram of the process $\tau^- \to \pi^- \pi^0 \nu_{\tau}$ with an intermediate $\rho$ meson.}
\label{tauInterm}
\end{figure}

The Feynman diagrams of the process $\tau^- \to \pi^- \pi^0 \nu_{\tau}$ without interaction in the final 
state are shown in Figs.~\ref{tauContact},~\ref{tauInterm}.
	
\begin{figure}[ht]
\center{\includegraphics[scale = 0.6]{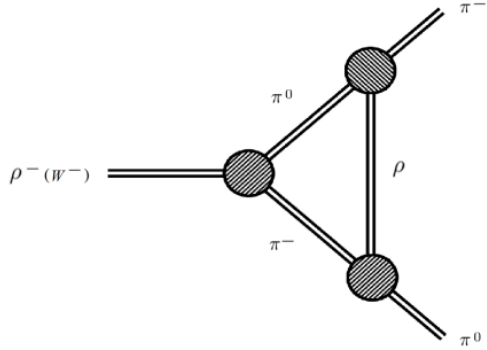}}
\caption{The pion interaction in the final state in the case of $\tau$ lepton decay.}
\label{rho_W_pi0pi}
\end{figure}

The meson triangle which is necessary for considering the pion interaction in the final state is shown 
in Fig.~\ref{rho_W_pi0pi}.

This triangle leads to the integral of the same form that in the case of the $e^{+}e^{-}$ annihilation process 
considered above. The amplitude of the $\tau$ lepton decay process takes the form
\begin{eqnarray}
M(\tau \to \pi \pi \nu) = -G_{F} V_{ud} \left[1 + \frac{s}{M_{\rho}^{2} - s - i \sqrt{s}\Gamma_{\rho}}\right] 
\left\{1 + g_{\rho}^{2}T_{\rho\pi\pi}\right\} L_{\mu}^{\mathrm{weak}} \left(p_{-} - p_{0}\right)^{\mu},
\end{eqnarray}
where $G_{F}$ is the Fermi constant, $V_{ud}$ is the element of the Cabibbo-Kobayashi-Maskawa matrix, 
$L_{\mu}^{\mathrm{weak}} = \bar{\nu}_{\tau}\gamma_{\mu}\left(1 - \gamma^{5}\right)\tau^{-}$ is the weak lepton current.
	
The first term in the first squared brackets describes the contact diagram, i.e., the diagram without an intermediate meson. 
The second term describes the diagram with an intermediate $\rho$ meson. The first term in the curly brackets corresponds 
to the diagram without final state pion interactions.
	
As a result, for the cut-off parameter determined in the previous Chapter, we obtain the partial width of this decay
\begin{eqnarray}
\mathrm{Br}(\tau \to \pi \pi \nu) = (25.1\pm 1.2)\%,
\end{eqnarray}
where the theoretical error $\sim \pm 5\%$ is estimated by means of comparison of the predictions of the chiral 
$SU(2)\times SU(2)$ NJL model with the experimental data of meson interaction including only light 
quarks~\cite{Volkov:1986zb}. The theoretical uncertainty is related to the effects of the chiral symmetry 
breaking and the pure leading in $1/N_c$ approximation of the model. 
	
The experimental partial width of this decay~\cite{Tanabashi:2018oca} is
\begin{eqnarray}
\mathrm{Br}(\tau \to \pi \pi \nu)_{\mathrm{exp}} = (25.49 \pm 0.09)\%.
\end{eqnarray}
	
Without considering final state interaction, the standard NJL model predicts for this decay 
the partial width of approximately 17\%. One can see that for this process, like in the case of the 
$e^{+}e^{-}$ annihilation into two pions, taking into account of the $\rho$ meson exchange between 
the pions is crucial to get a satisfactory agreement with experimental data.

\section{Conclusion} \label{sect:concl}
	
In this way, it was shown that in the considered processes the final state interactions play a significant role. 
This is due to the $\rho$ meson exchange between pions with a large coupling constant in the P-wave. 
Besides, the significance of the final state interaction contribution is caused by relative small pion energies 
and the closeness to the production threshold. The satisfactory agreement with the experimental 
data for both processes confirms the hypothesis of the vector current conservation.
We have to note that in the earlier articles~\cite{Volkov:2012tk,Volkov:2012uh} of one of the authors (M.K.V.)
the same processes were considered in a different version of the NJL model in a non-adequate manner. 
	
The correct description of the mesons interaction in the final state is impossible in the framework of the standard 
NJL model because it has been formulated in the lowest order on $1/N_{c}$. Whereas considering of the final state interaction 
of pions requires going beyond this approximation. This has been done in the present work. By using the known 
experimental data for the process $e^{+}e^{-} \to \pi^{+} \pi^{-}$ the value of the cut-off parameter of the meson 
momentum in the loop describing the pions interaction in the final state have been obtained. As a result, the decay 
$\tau^- \to \pi^- \pi^0 \nu_{\tau}$ has been described in a satisfactory agreement with the experimental data.
	
In the present work, we have considered only the domain of relatively small invariant masses of the pion pair 
near the $\rho$ meson peak because in this range the contribution of the final state interaction is significant. 
In the process of $e^{+}e^{-}$ annihilation into two pions, at the energies higher than 1~GeV, it is necessary to take 
into account the contributions of the diagrams with intermediate excited mesons, primarily, the $\rho(1450)$ meson. 
This can be done in the framework of the extended NJL model~\cite{Volkov:1996br,Volkov:1996fk,Volkov:1997dd}. 
However, at the considered energy region we effectively incorporate these contributions into the estimate of the 
standard NJL model theoretical uncertainty discussed above.


\begin{thebibliography}{99}

\bibitem{Nambu:1961tp} 
Y. Nambu and G. Jona-Lasinio, 
Phys. Rev. {\bf 122}, 345 (1961).
	
\bibitem{Ebert:1982pk}
D. Ebert and M.\,K. Volkov,
Z. Phys. C {\bf 16}, 205 (1983).
	
\bibitem{Volkov:1986zb}  
M.\,K. Volkov,  
Sov. J. Part. Nucl.  {\bf 17}, 186 (1986).
	
\bibitem{Ebert:1985kz} 
D. Ebert and H. Reinhardt, 
Nucl. Phys. B {\bf 271}, 188 (1986).
	
\bibitem{Volkov:1996br}
M.\,K. Volkov and C. Weiss,
Phys. Rev. D {\bf 56}, 221 (1997)
[arXiv:hep-ph/9608347 [hep-ph]].
	
\bibitem{Volkov:1996fk}
M.\,K. Volkov,
Phys. Atom. Nucl. {\bf 60}, 1920 (1997)
[arXiv:hep-ph/9612456 [hep-ph]].
	
\bibitem{Volkov:1997dd}
M.\,K. Volkov, D. Ebert and M. Nagy,
Int. J. Mod. Phys. A {\bf 13}, 5443 (1998)
[arXiv:hep-ph/9705334 [hep-ph]].
	
\bibitem{Volkov:2016umo} 
M.\,K.~Volkov and A.~B.~Arbuzov, 
Phys. Part. Nucl. {\bf 47}, 489 (2016).
	
\bibitem{Volkov:2017arr}
M.\,K.~Volkov and A.~B.~Arbuzov,
Phys. Usp. {\bf 60}, 643 (2017).
	
\bibitem{Volkov:2018zlr} 
M.\,K. Volkov and A.\,A. Pivovarov, 
JETP Lett. {\bf 108}, 347 (2018).
	
\bibitem{Volkov:2019cja} 
M.\,K. Volkov, A.\,A. Pivovarov and K. Nurlan, 
Eur. Phys. J. A {\bf 55}, 165 (2019).
	
\bibitem{Volkov:2019jug}
M.\,K. Volkov and A.\,A. Pivovarov, 
JETP Lett. {\bf 110}, 237 (2019).
	
\bibitem{Volkov:2019izp} 
M.\,K. Volkov, A.\,A. Pivovarov and K. Nurlan, 
Nucl. Phys. A {\bf 1000}, 121810 (2020).
	
\bibitem{Volkov:2019yli}
M.\,K. Volkov, A.\,A. Pivovarov and K. Nurlan,
Int. J. Mod. Phys. A {\bf 35}, 2050035 (2020)
[arXiv:1912.09812 [hep-ph]].
	
\bibitem{Tanabashi:2018oca} 
M. Tanabashi, K. Hagiwara, K. Hikasa et al. (Particle Data Group), 
Phys. Rev. D {\bf 98}, 030001 (2018).
	
\bibitem{Kuhn:1990ad}
J.\,H. Kuhn and A. Santamaria,
Z. Phys. C {\bf 48}, 445 (1990).
	
\bibitem{Bartos:2017oam}
E. Bartos, S. Dubnicka, A.\,Z. Dubnickova and H. Hayashii, 
Int. J. Mod. Phys. A {\bf 32}, 1750154 (2017).
	
\bibitem{Miranda:2018cpf} 
J.\,A. Miranda and P. Roig,
JHEP {\bf 11}, 038 (2018)
[arXiv:1806.09547 [hep-ph]].
	
\bibitem{Dai:2018thd}
L.\,R. Dai, R. Pavao, S. Sakai and E. Oset,
Eur. Phys. J. A {\bf 55}, 20 (2019).
	
\bibitem{Benayoun:2015hzf}
M. Benayoun,
EPJ Web Conf. {\bf 118}, 01001 (2016)
[arXiv:1511.01329 [hep-ph]].
	
\bibitem{Achasov:2005rg} 
M.\,N. Achasov, K.\,I. Beloborodov, A.\,V. Berdyugin et al.,
J. Exp. Theor. Phys. {\bf 101}, 1053 (2005).
	
\bibitem{Schael:2005am}
S. Schael et al. [ALEPH],
Phys. Rept. {\bf 421}, 191 (2005)
[arXiv:hep-ex/0506072 [hep-ex]].
	
\bibitem{Fujikawa:2008ma}
M. Fujikawa, H. Hayashii, S. Eidelman et al. [Belle],
Phys. Rev. D {\bf 78}, 072006 (2008)
[arXiv:0805.3773 [hep-ex]].
	
\bibitem{Actis:2010gg}
S. Actis, A. Arbuzov, G. Balossini et al. 
[Working Group on Radiative Corrections and Monte Carlo Generators for Low Energies],
Eur. Phys. J. C {\bf 66}, 585 (2010)
[arXiv:0912.0749 [hep-ph]].
	
\bibitem{Dax:2018rvs}
M. Dax, T. Isken and B. Kubis,
Eur. Phys. J. C {\bf 78}, 859 (2018)
[arXiv:1808.08957 [hep-ph]].
	
\bibitem{Weinberg:1978kz}
S. Weinberg,
Physica A {\bf 96}, 327 (1979).
	
\bibitem{Gasser:1983yg}
J. Gasser and H. Leutwyler,
Annals Phys. {\bf 158}, 142 (1984).
	
\bibitem{Truong:1991gv}
T.\,N. Truong,
Phys. Rev. Lett. {\bf 67}, 2260 (1991).
	
\bibitem{GomezNicola:2007qj}
A. Gomez Nicola, J.\,R. Pelaez and G. Rios,
Phys. Rev. D {\bf 77} (2008), 056006
[arXiv:0712.2763 [hep-ph]].
	
\bibitem{Ebert:1996pc}
D. Ebert, T. Feldmann and M.\,K. Volkov,
Int. J. Mod. Phys. A {\bf 12}, 4399 (1997).
	
\bibitem{Volkov:2009pc}
M.\,K. Volkov, E.\,A. Kuraev and Y.\,M. Bystritskiy,
Central Eur. J. Phys. {\bf 8}, 580 (2010).
	
\bibitem{Volkov:2012tk}
M.~K.~Volkov and D.~G.~Kostunin,
Phys. Rev. C \textbf{86} (2012), 025202
[arXiv:1204.1455 [hep-ph]].

\bibitem{Volkov:2012uh}
M.~K.~Volkov and D.~G.~Kostunin,
Phys. Part. Nucl. Lett. \textbf{10} (2013), 7-10
[arXiv:1202.0506 [hep-ph]].
	
\end{thebibliography}
\end{document}